\documentclass[prl,twocolumn,showpacs,superscriptaddress,preprintnumbers,amsmath,amssymb]{revtex4-1}
\usepackage{graphicx}
\usepackage{dcolumn}
\usepackage{bm}
\usepackage[colorlinks=true,linkcolor=blue]{hyperref}
\usepackage{CJK,CJKnumb,CJKulem}
\begin{document}
\begin{CJK*}{GBK}{com}
\title{Hyperuniform Mixing of Binary Active Spinners}
\author{Rui Liu}
\email{lr@iphy.ac.cn}
\affiliation{Beijing National Laboratory for Condensed Matter Physics and CAS Key Laboratory of Soft Matter Physics, Institute of Physics, Chinese Academy of Sciences, Beijing 100190, China}
\author{Mingcheng Yang}
\affiliation{Beijing National Laboratory for Condensed Matter Physics and CAS Key Laboratory of Soft Matter Physics, Institute of Physics, Chinese Academy of Sciences, Beijing 100190, China}
\affiliation{School of Physical Sciences, University of Chinese Academy of Sciences, Beijing 100049, China}
\author{Ke Chen}
\affiliation{Beijing National Laboratory for Condensed Matter Physics and CAS Key Laboratory of Soft Matter Physics, Institute of Physics, Chinese Academy of Sciences, Beijing 100190, China}
\affiliation{School of Physical Sciences, University of Chinese Academy of Sciences, Beijing 100049, China}
\date{\today}
\begin{abstract}
Spinner mixtures consisting of both clockwise and counterclockwise self-spinning particles are often expected to phase separate. However, we demonstrate that such a demixing is absent for dimer (or rod-like) spinners. These particles always mix, even in a globally-hyperuniform way, with the total structure factor $S(q\to 0)\sim q^{\alpha}\,(\alpha>0)$. This global hyperuniformity can be enhanced or weakened by changes in the driving torques or the particle density 
in various ways. The corresponding microscopic mechanism is attributed to the competition between a dynamical heterocoordination effect and effective like-particle attractions. Critical scaling for the absorbing state transition of the system is also found to persist, with a significant shift in its critical point observed. The system can be further thermalized, by the driving torques or through thermostating, into an ideal solution with identical partial radial distribution functions, which denys the possibility of being multi-hyperuniform. A simply-extented coupled density-oscillator theory explains why the system can not be multi-hyperuniform, but can have a global hyperuniformity with the scaling exponent $\alpha$ approaching $2$. Such a hyperuniform mixing provides a way to regulate the topological boundary flows of this chiral system, and this mixing regulation is found to barely affect the bulk density fluctuations and even preserve the localization of the flows and the bulk hyperuniformity.
\end{abstract}
\maketitle
\end{CJK*}

\section{Introduction}
Disordered hyperuniformity is an exotic property of matter, which indicate that the structure is isotropic as liquids but suppress long-wavelength density fluctuations as crystals \cite{Torquato2018,Torquato2003,Torquato2016}. The concept has been extended to binary or multi-component systems, in which two-phase hyperuniformity \cite{Chen2018,Ma2020}, global hyperuniformity \cite{Lomba2017} and multi-hyperuniformity \cite{Jiao2014,Lomba2018} have been extensively investigated. Hyperuniformity properties are also known to persist in the active or fluidic states of single-component nonequilibrium/active systems \cite{Hexner2017,Lei2019,Liu2023,Zhang2022}. 

Binary or multi-component fluids may mix. Fluids mixing is a fundamental process in nature and industry, and a common problem in this process is whether a uniform mixture of different components can be obtained. External disturbing/driving, such as stirring, is usually employed to ensure that the liquid can be uniformly mixed. Active matter is driven by its internal energy sources, thus would be usually more likely to mix. Whether or how hyperuniformity would persist for active fluidic mixtures remains to be explored. A recent study on a robot mixture with progammed nonreciprocal interactions \cite{Chen2024} shows that hyperuniformity may exist at least in its critical absorbing state. 

Active spinner systems, which consist of self-spinning particles, have been widely studied in recent years. They are known to exhibit a variety of interesting behaviors, such as phase separation \cite{Nguyen2014,Scholz2018,Yeo2015}, jamming \cite{vanZuiden2016}, and topological effects \cite{Dasbiswas2018,Souslov2019,Yang2021,Liu2020}. It is known that spinner mixtures consisting of both clockwise and counterclockwise self-spinning particles tend to phase separate due to effective like-particle attractions \cite{Nguyen2014,Scholz2018}. However, in this paper, we demonstrate that such a demixing behavior is absent for dimer (or rod-like) spinners. These particles always mix, even in a hyperuniform way. 

Through numerical simulations, we reveal that hyperuniformity typically persists in a global way for such a binary spinner fluid. The competition between dynamical heterocoordination and effective like-particle attraction can affect the long-wavelength scaling law of the total structure factor of the system. The heterocoordination effect may also cause a significant shift in the critical point of the absorbing transition of the system. Then, we show that the increase in driving torques may enhance the global hyperuniformity and that in density may somehow weaken it, leading to a collapsing long wavelength behavior in the total structure factor. Further, the mixrure is shown to be thermalized into an ideal solution through either increasing the driving torque or particularly by thermostating. We also extend the density oscillator model by Lei \& Ni \cite{Lei2019}, simply through introducing some linear couplings, to this binary system, which explains why the system can not be multi-hyperuniform, but can have a global hyperuniformity with the scaling exponent $\alpha$ approaching $2$. Finally, we show that such a mixing will not affect bulk density fluctuations, and may even preserve localization and bulk hyperuniformity in the presence of topological boundary flows, which thus provides a intriguing way to regulate the robust flows.

\section{Simulation}
We simulate a two-dimensional (2D) active spinner system as that in our previous study \cite{Liu2023}. Each spinner is a dimer consisting of two spherical monomers bonded with a fixed length $\sigma=1$. Each monomer has a mass $m=1$ and each dimer has a moment of inertia $I=\frac{1}{2}m\sigma^2$. The monomers from different dimers interact with each other through a Weeks-Chandler-Andersen potential $U(r_{ij})=4\epsilon[(\frac{\sigma}{r_{ij}})^{6}-(\frac{\sigma}{r_{ij}})^{12}]+\epsilon$, when their separation $r_{ij}$ is smaller than a cutoff distance $r_c\equiv2^{1/6}\sigma$. With the total effect of all such pair interactions denoted by $U(t)$, the dynamics of any dimer $i$ is governed by 
\begin{gather}
2m\ddot{\bf r}_i=-2\gamma_t\dot{\bf r}_i-\nabla_i U(t)+\bm{\xi}_i(t),\\
I\ddot{\theta}_i=\pm \tau-\gamma_s\dot{\theta}_i-\partial_{\theta_i}U(t)+\zeta_i(t).
\end{gather}

Assuming $\epsilon=1$ energy unit, and taking $\hat{t}=\sqrt{m\sigma^2/\epsilon}$ to be the time unit, we always set the translational frictional coefficient $\gamma_t=m/\hat{t}$ and the rotational counterpart $\gamma_s=I/\hat{t}$ in our simulation. The driving torques $\pm \tau$ are respectively applied to make the dimers spin in the clockwise and counterclockwise directions. The timestep of the simulation is adjusted accordingly for different driving torques and thermostating temperatures. All simulations run for no less than $3\times 10^7$ time steps, to ensure the system reaches a steady state and good statistical results can be obtained. $\bm{\xi}(t)$ and $\zeta(t)$ are respectively the stochastic force and torque due to thermal fluctuations: $\langle \xi_i(t)\xi_j(t')\rangle=2\gamma_t k_B T \delta_{ij}\delta(t-t')$, $\langle \zeta_i(t)\zeta_j(t')\rangle=2\gamma_s k_B T \delta_{ij}\delta(t-t')$. Typically, a system with $N$ dimers in a square box of size $L$ is simulated, and the number density is evaluated as $\rho=N/L^2$, or $\phi=N\sigma^2/L^2$ in a dimensionless way. Periodic boundary conditions are always applied for all box boundaries. Similar measurements are adopted for the system simulated in a disc container to demonstrate the regulation of topological boundary flows, where a smooth spherical wall is used to confine the system. 

\section{Theory}
The dimers are driven through torques acting on each monomers, which thus preserve the center of mass conservation (COMC). COMC is crucial for density hyperuniformity \cite{Hexner2017,DeLuca2024}, and this provides a general explanation why spinner systems posess hyperuniform properties in various manners \cite{Lei2019,Liu2023}. By neglecting the spin-orbit coupling, and encoding the driving effect into the kinetic temperature $T_k$, a generic density oscillator theory for such active fluids is given by Lei \& Ni \cite{Lei2019}:
\begin{gather}
    \frac{\partial^2 \delta \rho}{\partial t^2}=-\Gamma_{q}\frac{\partial \delta \rho}{\partial t}-Dq^2\delta \rho + q^2\sigma^r+q \sigma^t \label{denosc},
\end{gather}
where $\delta\rho$ is the density fluctuation, $q$ is the modulus of the wave vector $\mathbf{q}$, $\Gamma_{q}$ desribes the total effect of both the substrate friction and kinematic viscosities, $D$ describes the diffusional effect, $\sigma^r$ denotes the longitudinal term of the collisional noise. Transverse modes are ignored since they are irrelevant to the density fluctuations. Additionally, a longitudinal thermal noise term $\sigma^t$ is added here. 

A simple extension to the binary system would be assuming linear coupling between the density fluctuations of the two species: $\delta \rho=(\delta\rho_1,\delta\rho_2)^{\mathrm{T}}$. Thus the dynamical coefficients $\Gamma_{q}$ and $D$ become matrices:
\begin{gather}
    \Gamma_{q}=\begin{bmatrix} \gamma+\eta q^2 & \chi \\ \chi & \gamma+\eta q^2 \end{bmatrix},\quad D=\begin{bmatrix} c_s^2 & \beta \\ \beta & c_s^2 \end{bmatrix},
\end{gather}
where $\gamma=\gamma_t/m$ is the reduced frictional coefficient, $\eta$ is the longitudinal viscosity, $c_s$ is the sound speed, $\chi$ describes the inter-species momentum transfer as a frictional term, and $\beta$ measures the effect of inter-species pressure on diffusion. 

Obviously, each species by itself does not preserve COMC. Thus the collisional noise $\sigma^r$ can not be naively decomposed into two surfacial terms. We assume, for each species, the noise can be decomposed into a gradient term and a thermal-like noise. Thus we have
\begin{gather}
    q^2\sigma^r\longrightarrow (q^2\sigma^{r}_1+\mathbf{q}\cdot \bm{\delta\sigma}^{r},q^2\sigma^{r}_2-\mathbf{q}\cdot \bm{\delta\sigma}^{r})^{\mathrm{T}}.
\end{gather}

In this way, we have the separated COMC terms of collisional noises $\tilde{\sigma}^r=(\sigma_1^r, \sigma_2^r)^{\mathrm{T}}$ and effective longitudinal thermal noises $\tilde{\sigma}^t=(\sigma_1^t+\delta\sigma^r,\sigma_2^t-\delta\sigma^r)^{\mathrm{T}}\equiv(\sigma_1^t+\delta\sigma_1^r,\sigma_2^t+\delta\sigma_2^r)^{\mathrm{T}}$. By performing a temporal Fourier transform, Eq.\ref{denosc} can be rewriten as:
\begin{gather}
    (-\omega^2I-i\omega\Gamma_{q}+ Dq^2)\delta\rho(q,\omega)=q^2\tilde{\sigma}^r+q\tilde{\sigma}^t.
\end{gather}

Assuming all random terms are both spatially and temporally white: $\langle\sigma_\mu^r\sigma_\nu^r\rangle(\mathbf{q},\omega)=a(T_k)\rho\sqrt{x_\mu x_\nu}\delta_{\mu\nu}$, $\langle\delta\sigma_\mu^r\delta\sigma_\nu^r\rangle(\mathbf{q},\omega)=b(T_k)\rho\sqrt{x_\mu x_\nu}\epsilon_{\mu\nu}$, and $\langle \sigma_\mu^t\sigma_\nu^t\rangle(\mathbf{q},\omega)=c(T)\rho\sqrt{x_\mu x_\nu}\delta_{\mu\nu}$ ($x_{\mu,\nu}$ are respectively the concentrations of species $\mu,\,\nu$; $T$ is the temperature of the thermostat and the driven system has a non-zero kinetic temperature $T_k$ even at $T=0$; $\epsilon_{\mu\nu}=1$ for $\mu=\nu$, and $-1$ for $\mu\ne\nu$), one obtains the following results for an equimolar system ($x_1=x_2=1/2$):
\begin{gather*}
    S_{\mu\nu}(q,\omega)=\langle\delta\rho_{\mu}\delta\rho^{*}_{\nu}\rangle\\=\sum\limits_{\kappa\lambda}M_{\mu\kappa}M^*_{\nu\lambda}[q^4\langle \sigma_\kappa^r\sigma_\lambda^r\rangle + q^2 \langle\tilde{\sigma}_\kappa^t \tilde{\sigma}_\lambda^t\rangle]\\
    = \frac{1}{2}(aq^4+cq^2)\rho\sum\limits_{\lambda}M_{\mu\lambda}M^*_{\nu\lambda}+\frac{1}{2}bq^2\rho\sum\limits_{\kappa\lambda}M_{\mu\kappa}M^*_{\nu\lambda}\epsilon_{\kappa\lambda},
\end{gather*}
where $\langle\cdot \rangle$ denotes an ensemble average, $M=(-\omega^2I-i\omega\Gamma_{q}+ Dq^2)^{-1}$. Due to the symmetry of $M$, all calculation results will reduce to matrix elements of $K=MM^*$:
\begin{align*}
    S_{11}(q,\omega)=& S_{22}(q,\omega) =\\ & \frac{\rho}{2}[(aq^4+cq^2)K_{11}+bq^2(K_{11}-K_{12})], \\
    S_{12}(q,\omega)=& S_{21}(q,\omega) =\\ & \frac{\rho}{2}[(aq^4+cq^2)K_{12}-bq^2(K_{11}-K_{12})].
\end{align*}
Then the total structure factor is given by:
\begin{align*}
    S(q,\omega)&=\sum\limits_{\mu\nu}S_{\mu\nu}(q,\omega)\\&=\rho(aq^4+cq^2)(K_{11}+K_{12}).
\end{align*}
The $bq^2$ terms simply cancel out, and the $cq^2$ terms will not present for the case without thermal noise. By integrating with $\omega$, one obtains:
\begin{gather*}
    \int_{-\infty}^{\infty}(K_{11}\pm K_{12})d\omega =\frac{\pi}{q^2(c_s^2\pm \beta)(\gamma+\eta q^2\pm \chi)}.
\end{gather*}
For $T=0$, we simply have $S(q\to 0)\sim q^2$:
\begin{gather}
    S(q)=\int_{-\infty}^{\infty}S(q,\omega)d\omega=\frac{\pi\rho aq^2}{(c_s^2+\beta)(\gamma+\eta q^2+\chi)}.
\end{gather}

Thus, theoretically, hyperuniformity may persist in such a binary fluid, keeping the scaling exponent $\alpha=2$. However, the inter-species and inner-species spiner-spinner interactions are generally different, thus there are pairing noise which cause local concentration fluctuations. The pairing noise preseves COMC only globally, but not locally. This may lead to $q$-dependencies in the above correlation functions of the random terms, which would further cause a decrease in the scaling exponent $\alpha$. Thus a weaker hyperuniformity characterized by a smaller $\alpha$ is acceptable.


One may expect the binary system to be multi-hyperuniform. However, we have $S_{11}\sim aq^2+b$ with a nonvanishing $q$-independent term $b$, which indicates that the subsystem consitituted by one of the two species is inevitably non-hyperuniform. Thus we do not have a multi-hyperuniformity in general. Nevertheless, the global hyperuniformity with a scaling exponent $\alpha$ approaching $2$ can be expected.

\section{Dynamical Heterocoordination and Global Hyperuniformity}
\begin{figure*}
    \includegraphics[height=0.3\linewidth]{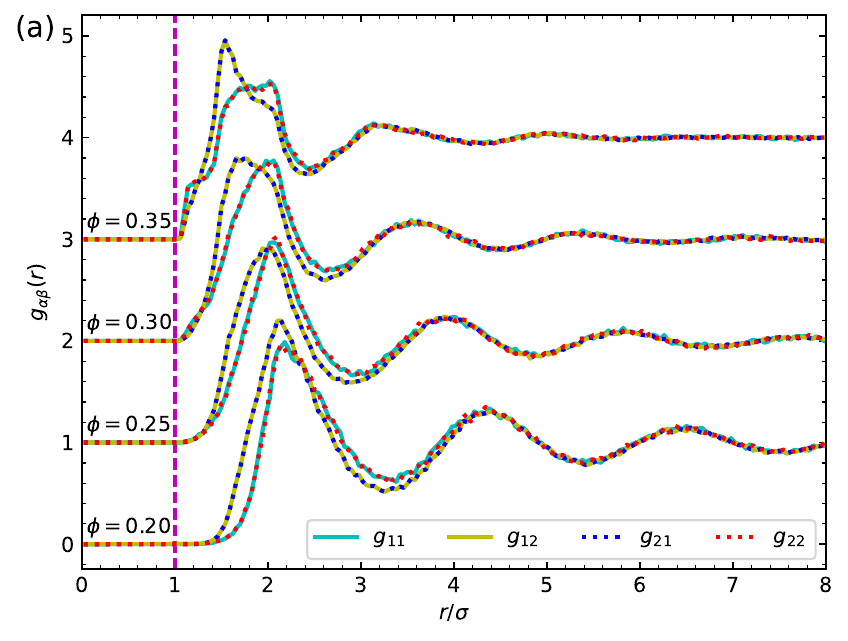}
    \includegraphics[height=0.3\linewidth]{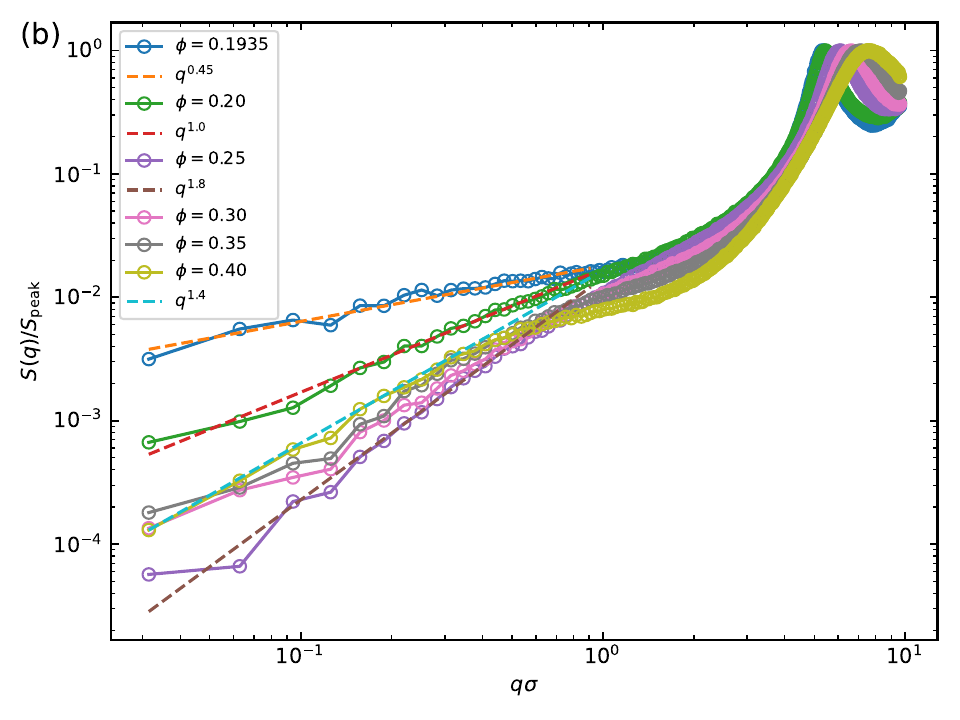}
    \caption{\label{figT0} The binary active spinner system at $\tau=1,\,T=0,\,L=200\sigma$, and varying $\phi=0.20, 0.25, 0.30, 0.35$: (a) partial radial distribution functions $g_{\alpha\beta}(r)$ (vertically shifted by 1 for each $\phi$, which is adopted below without further explainations); (b) the total structure factor, with $S(q\to 0)\sim q^{\alpha}$.}
\end{figure*}

Numerically, we first investigate such a spinner system without thermal noise, i.e. $T=0$. Fig.\ref{figT0} shows our simulation results of the system at different densities. We observe an effective unlike-particle attraction for all the cases: the cross terms $g_{\mu\mu}(r)$ ($\mu=1,2$) of the partial radial distribution functions surpass the diagonal terms $g_{\mu\nu}(r)$ ($\mu\ne \nu$) just before/at the major peak, as shown in Fig.\ref{figT0}(a). This is similar to that of the negatively-nonadditive hard-disk plasmas \cite{Lomba2017}, which show a heterocoordination effect and do not have a demixing transition. These effects are demonstrated to be compatible with a global hyperuniformity for the binary mixture.

Demixing is also found to be absent here for our dimer (or generically rod-like) spinners, and we expect a similar global hyperuniformity to be observed as that in negatively-nonadditive hard-disk plasmas. We calculate the total structure factor \cite{Lomba2017,Lomba2018} for this system:
\begin{gather}
    S(q)=1+\rho \sum\limits_{\mu\nu}x_\mu x_\nu \mathcal{F}[g_{\mu\nu}(r)-1](q) \nonumber \\=1+\rho \sum\limits_{\mu\nu}x_\mu x_\nu h_{\mu\nu}(q)
\end{gather}
where $\mathcal{F}[\cdot]$ represents a Fourier transform, $h(q)\equiv \mathcal{F}[h(r)](q)=\mathcal{F}[g(r)-1](q)$. The corresponding ressults are shown in Fig.\ref{figT0}(b). 

Generally, we have the power-law scaling in the total structure factor $S(q\to 0)\sim q^{\alpha}$ with $\alpha \gtrsim 1$ for all $\phi\ge 0.20$, which indicates a strong global hyperuniformity for all densities in the active fluidic regime. The system stays in an absorbing state below $\phi_c \approx 0.19$ [which has the critical scaling exponent $\alpha_c\approx 0.45$ as shown in Fig.\ref{figT0}(b)], and becomes a mixed active fluid above this critical point. Compared with Lei \& Ni's result that $\phi_c < 0.15$ \cite{Lei2019}, we have an observable shift in the critical density, due to the dynamical heterocoordination effect. Another significant difference in this binary system is that it does not exhibit the same scaling $S(q)\sim q^2$ beyond the critical point, i.e. in the active state, which holds for single-component systems \cite{Hexner2017,Lei2019}. 

We have a maximum exponent $\alpha\approx 1.8$ at about $\phi=0.25$, which is close to the value $\alpha=2$ for single-component spinners. We assume this correponds to the predicted results $S(q)\sim q^2$ in the theory section. The exponent first increases from the critical value $0.45$ to $1.8$ and then decreases to a relative lower value $\alpha=1.4$, as the density increases, giving a varying scaling law at long wavelengths. We argue that the varying scaling law and the deviation from $\alpha =2$ are due to the competetion between dynamical heterocoordination and some effective like-particle attraction. We observe that for a denser system at $\phi=0.35$, $g_{11}(r)$ becomes higher than $g_{12}(r)$ just before the major peak [Fig.\ref{figT0}(a)], which may indicate an effective like-particle attraction. The dynamical heterocoordination effect is likely to enhance mixing, while the effective like-particle attraction tries to induce a phase separation. The competition between these two effects cause the so-called pairing noise, which does not preserve COMC locally. Thus, one observes such a non-monotonic change in the scaling exponent $\alpha$.

Previously, the effective like-particle attractions are adopted to expain how binary spinners phase separate \cite{Nguyen2014,Scholz2018}. The microscopic picture for such an attraction is depicted as: pairs of like particles stay together relatively longer than unlike pairs during collisions. However, for the dimmer or rod shape spinners here, we argue that like pairs have large relative tangential velocities during collisions, a system minimizing its dissipation would avoid like pairing in its steady states. Or equivalently, unlike dimers with proper phase differences can stay closer into each other's sweeping range, which can be effectively treated as an attraction. All these effects are subtle. What favors unlike pairing may be roughly compensated by what else favors like pairing. This may finally lead to an ideal mixing with identical radial distribution functions for the two species (see below).

\section{Enhanced Hyperuniformity and Self-Thermalization}
\begin{figure*}
    \includegraphics[height=0.3\linewidth]{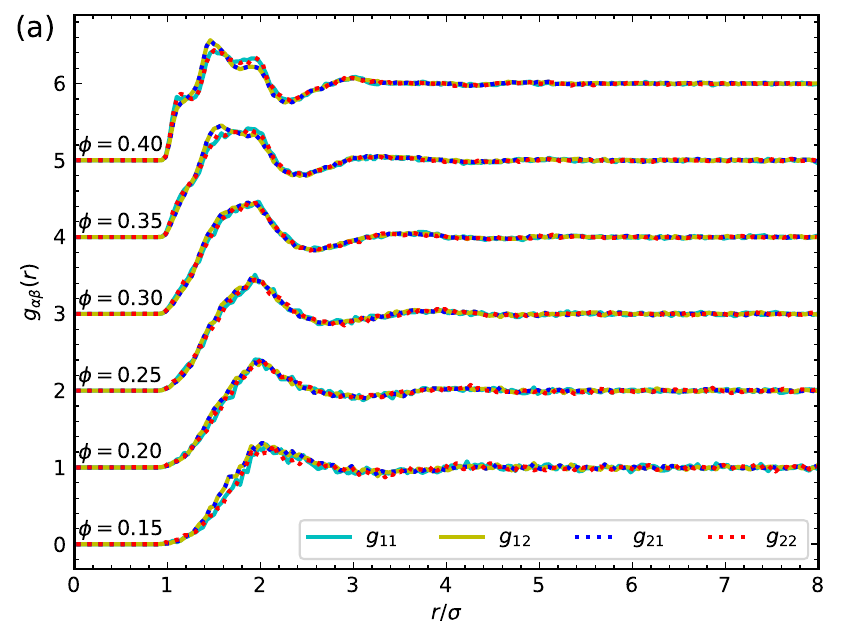}
    \includegraphics[height=0.3\linewidth]{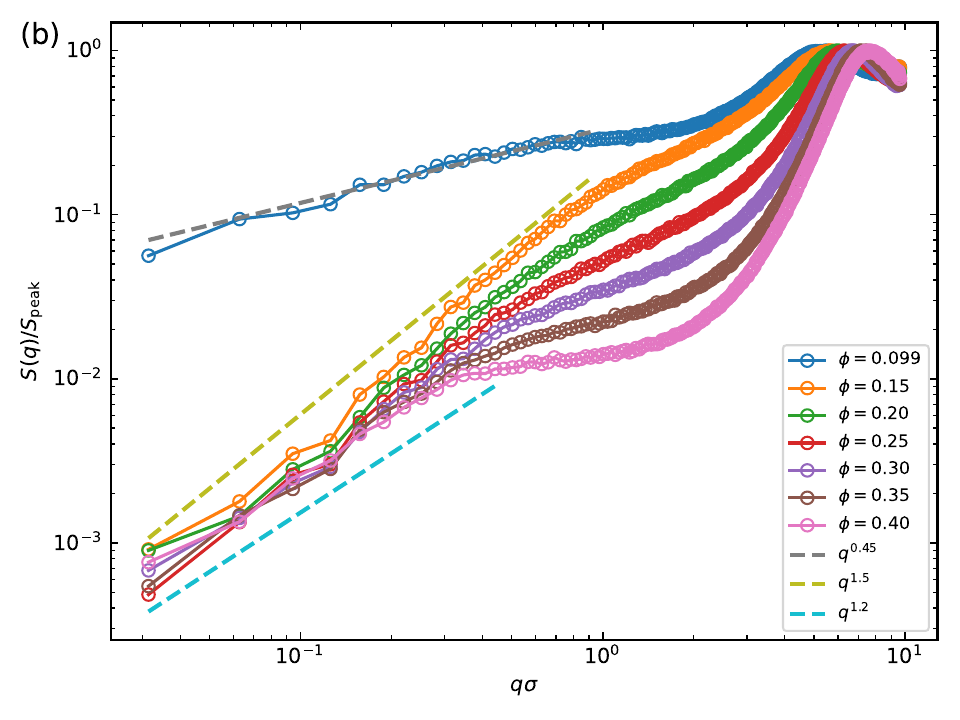}
    \caption{\label{figtorq5} The binary active spinner system at $\tau=5,\,T=0,\,L=200\sigma$: (a) the partial radial distribution functions $g_{\alpha\beta}(r)$; (b) the total structure factor $S(q)$; dashed lines show some asymptotic scalings at long wavelengths.}
\end{figure*}

The driving torques of the spinners would enhance the global hyperuniformity at low densities. An apparent example would be the system at density $\phi=0.15$, as shown in Fig.\ref{figtorq5}. At $\tau=1$, the system is below the critical point of the absorbing transition, which is thus intrinsically non-hyperuniform. However it becomes strongly hyperuniform with the scaling exponent $\alpha \gtrsim 1.5$ at $\tau=5$, as shown in Fig.\ref{figtorq5}(b). The critical density of the transition decreases to an even lower value $\phi_c\approx 0.099$ [the data with $q^{0.45}$ scaling as a guide of eye in Fig.\ref{figT0}(b)]. For the system at $\phi=0.2$, which is in the active state for both $\tau=1$ and $5$, a significant promotion in the scaling exponent $\alpha$ from the critical value $0.45$ to about $1.5$ is also observed, compared with the data shown in Fig.\ref{figT0}(b), 

However, for larger densities, high torques seem to be not beneficial to hyperuniformity. As can be seen from Fig.\ref{figtorq5}(b), the scaling exponent $\alpha$ for $\phi=0.35$ or $0.4$ is just about $1.2$, which is lower than that of $\phi=0.2$, or even the corresponding value of its own at $\tau=1$ in Fig.\ref{figT0}(b). This weakening in hyperuniformity is due to the effective like-particle attraction, identified by $g_{11}(r)>g_{12}(r)$ just beyond $r=1\sigma$ as shown in Fig.\ref{figtorq5}(a). This effect dominates only at nearer separations of the spinners, which thus usually requires a larger density for the system.

Moreover, high torques seem to make the long wavelength (small $q$) behaviors of $S(q)$ at different $\phi$ collapse, but distingush the density differences in the intermediate range of $q$. The scaling exponent increase fast from the citical value $\alpha_c=0.45$ to about $1.5$, and stays in roughly the range $(1.2, 1.5)$ for a wide range of densities $\phi \gtrsim 0.15$, as shown in Fig.\ref{figtorq5}(b). The collapse is due to the joint effect of torque-enhancing and density-weakening effects mentioned above. The intermediate range at the order of $q\sigma\sim 1$ coreponds to the length scale of spinner-spinner interactions, and structures at these length scales may be significantly changed by the torques.

Accompanied with the enhancement of hyperuniformity, we observe an identification of all partial radial distribution functions $g_{\alpha\beta}(r)$ of the system, for densities $\phi<0.35$ [Fig.\ref{figtorq5}(a)]. All $g_{\alpha\beta}(r)$ become more and more identical as the density increases from $0.15$ to $0.30$. This is obviously due to the increased probability of spinner-spinner interactions. Larger torques would heat the system up more easily and thus also enhance the spinner-spinner interactions. Identical partial radial distributions correpond to an ideal solution described by the substitutional model of Faber \& Ziman \cite{Faber1965}, which has a constant concentration-concentration structure factor $S_{cc}(q)$. Thus the system can not be multi-hyperuniform, which requires both vanishing $S(q)$ and $S_{cc}(q)$ as $q$ approaches zero. However, the system can still be globally hyperuniform as discussed above, since we do not introduce any thermal noise ($T=0$) which does not abide by COMC. The current thermalization only causes a slight decrease in the scaling exponent $\alpha$, as the density increases. As such a thermalization is much easier to be achieved in the presence of thermal noise, we will discussed it further below. 

\section{Thermalization and Ideal Solution}
\begin{figure*}
    \includegraphics[height=0.3\linewidth]{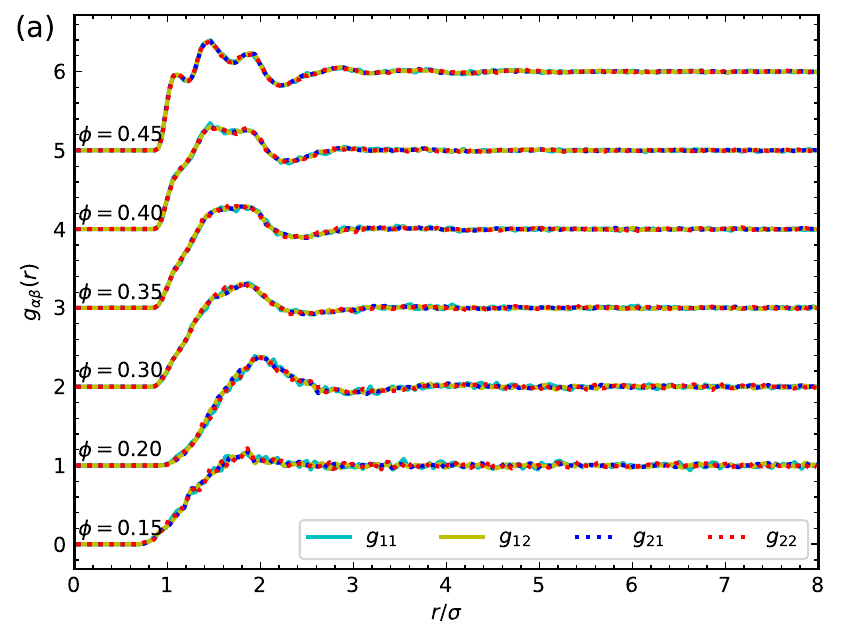}
    \includegraphics[height=0.3\linewidth]{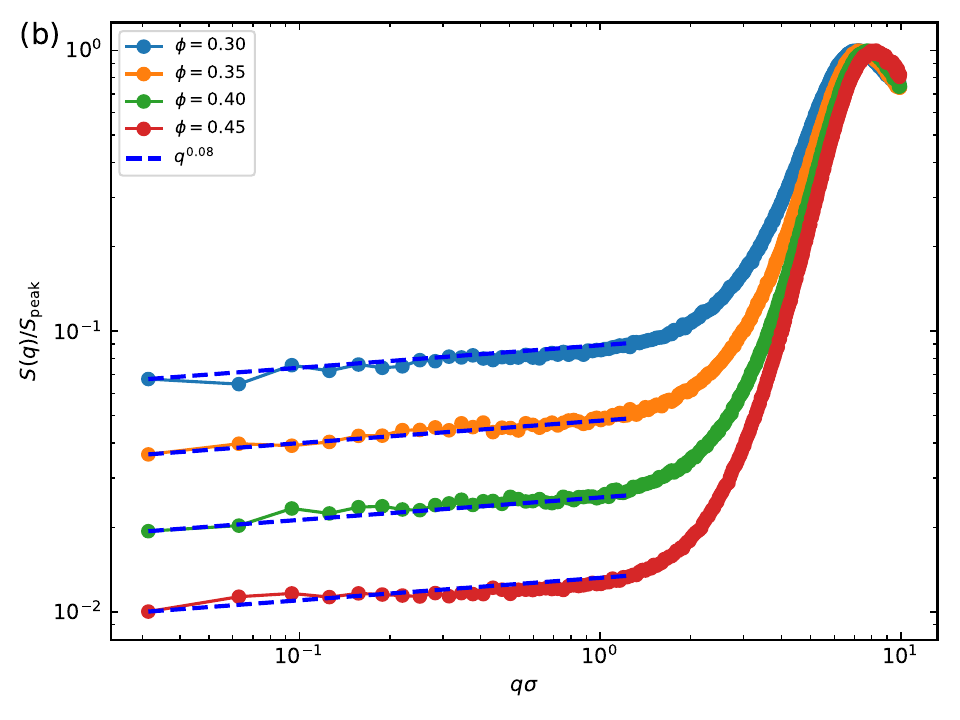}
    \caption{\label{figT1} Thermalization of the binary spinner system at $\tau=1,\,T=1,\,L=200\sigma$: (a) identical partial radial distribution functions $g_{\alpha\beta}(r)$ at various densities; (b) the total structure factor $S(q)$, with a drastic decrease in $S(q\to 0)$ as the density increases; dashed lines show a $q^{0.08}$ scaling.}
\end{figure*}

Thermal noise is known to weaken or destroy hyperuniformities \cite{Lei2019b}. We also investigate the thermal effects on the global hyperuniformity of this binary mixture. We observe a notable thermalization characteristic for a wide range of densities:
\begin{gather}
    g_{11}(r)=g_{22}(r)=g_{12}(r)=g_{21}(r).
\end{gather}
Namely, we have again identical partial radial distribution functions [as shown in Fig.\ref{figT1}(a)], which correpond to an ideal solution with constant concentration-concentration fluctuations \cite{Lomba2017,Lomba2018,Bhatia1970} at all modes, i.e.:
\begin{gather}
    S_{cc}(q)=x_1x_2\left[1+\rho x_1x_2\sum\limits_{\mu\nu}\epsilon_{\mu\nu}h_{\mu\nu}(q)\right]=x_1x_2.
\end{gather}
For an equimolar binary system, one has $S_{cc}(q)=0.25$.

Thus, as mentioned above, there is no chance for the system to become multi-hyperuniform. We still calculate the total structure factor $S(q)$ for the system, which is shown in Fig.\ref{figT1}(b). We have a robust scaling relation $S(q) \sim  q^{0.08}$ at small $q$, though the universal exponent is rather small. This could be the remnant of the above non-thermal hyperuniformity, and one may conclude that such a thermalized system is rather weakly hyperuniform or just nonhyperuniform in the global sense. 

However, we still observe a drastic decrease in $S(q\to 0)$ as the density of the system increases. This is due to the fact the system starts to pack for increased density. Such a packing effect would lead to a jamming type hyperuniformity \cite{Zachary2011,Hexner2018} or spinners' nonhyperuniformity with lowly-confined density fluctuations \cite{Liu2023}. In either case, the long-wavelength behavior $S(q\to 0)$ should be decreased, as has been observed.

\section{Preserved Hyperuniformity in Mixing Regulation}
The chiral system of single-component spinners is known to exhibit robust topological boundary flows \cite{Dasbiswas2018,Souslov2019}. Usually, it would be a challenge to regulate such a flow without altering the corresponding carrier density. Since binary spinners can mix hyperuniformly, we may utilize such a feature to tune the topological boundary flow by simply adjusting the concentrations of the two species in this system \cite{topub}. Such a regulation would be preferable in experimental systems, especially for sealed samples where density of particles can be hardly changed.

\begin{figure*}
    \includegraphics[height=0.3\linewidth]{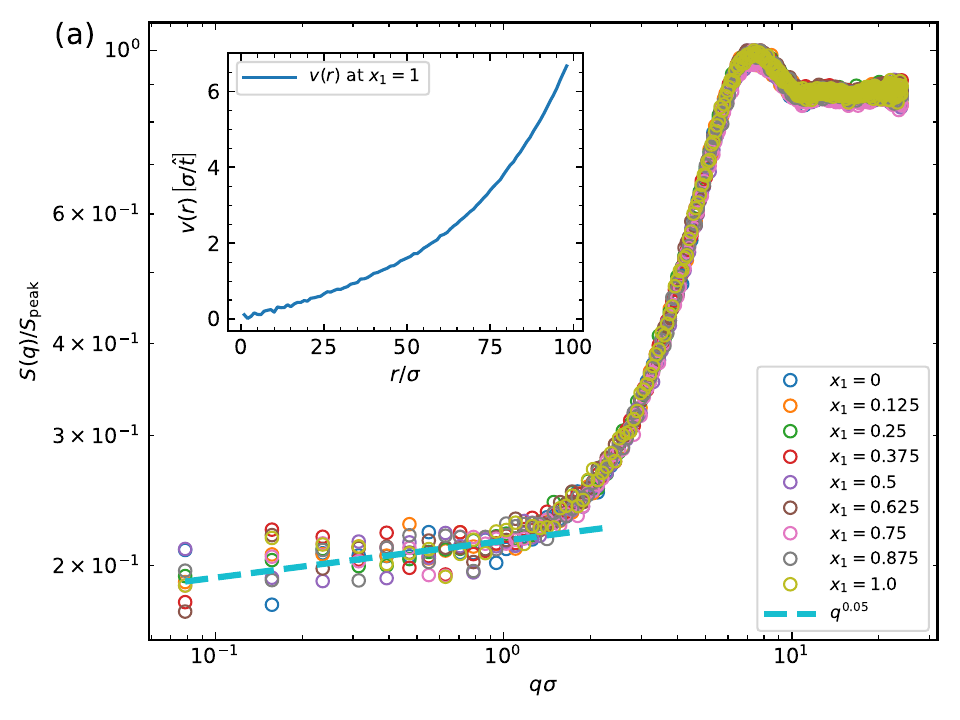}
    \includegraphics[height=0.3\linewidth]{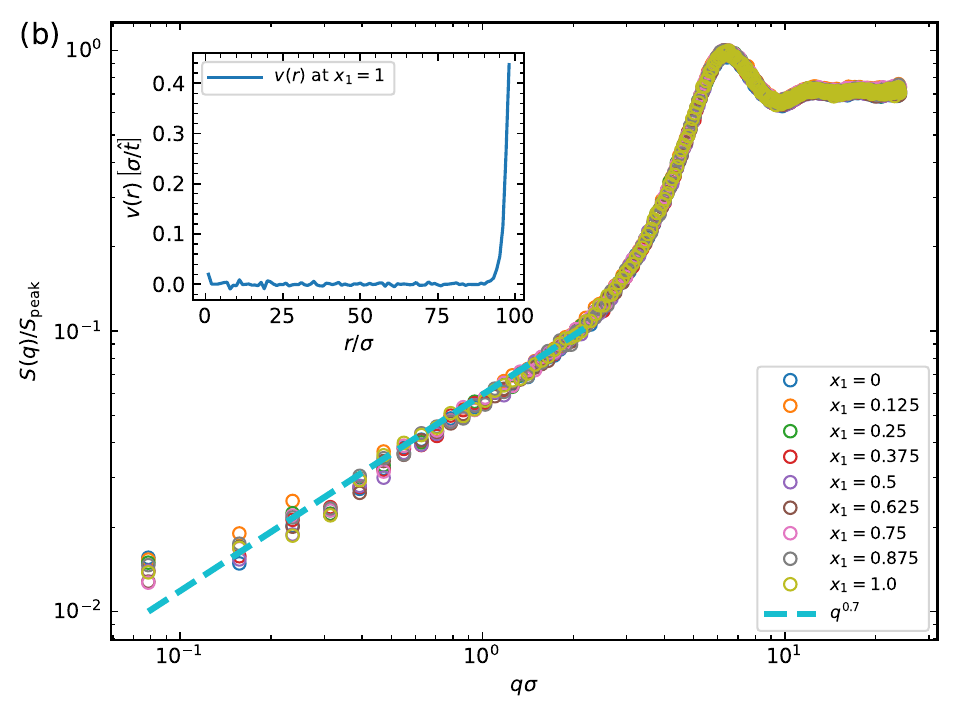}
    \caption{\label{figflow} Binary spinners in a disc of radius $R=100\sigma$ simulated with $T=0,\,\tau=5,\,\phi\approx 0.255$: (a) Nearly collapsed total structure factor $S(q)$ for various $x_1$ in the presence of weakly-localized boundary flows, the inset plots the radial flow profile $v(r)$ at $x_1=1$; (b) Nearly collapsed total structure factor $S(q)$ for various $x_1$ in the presence of strongly-localized boundary flows, the inset plots the flow profile at $x_1=1$. All $S(q)$ measurements are performed in a centered square subarea of size $L_s=0.8R$.}
\end{figure*}

Here, we would focus on the regulation effect on the density fluctuations in the bulk. Fig.\ref{figflow}(a) shows the total structure factor $S(q)$ corresponds to a system with weakly-localized boundary flow in a disc of size $R=100\sigma$, where the flow profile extends to the center of the disc (see the inset). The measurement of $S(q)$ is performed in a square subarea of size $L_s=0.8R$ at the center of the disc. The subsystem does not preserve COMC globally, due to the in- and out-flows of particles. For the existence of the obivous flow, the whole system possesses no obvious hyperuniformity features. While the regulation is effective for the whole flow field, we observe an preservation in the density fluctuations of the bulk: the total structure factor $S(q)$ is nearly collapsed for different concentrations in the full range $[0,1]$, which indicates that the density fluctuations in the bulk are not significantly altered by the regulation.

When the topological boundary flow is strongly localized, with an inner zero velocity field, the regulation will only affect the boundary flow, and abides by localization. No abvious flow in the bulk admits a global hyperuniformity of the mixture, though COMC is not globally preserved. In Fig.\ref{figflow}(b), we show the total structure factor $S(q)$ for the same system with a strongly localized boundary flow (see the inset). The total structure factor is again nearly collapsed for different concentrations in the full range $[0,1]$, but accompanied with an obvious hyperuniformity feature $S(q\to 0)\sim q^{0.7}$. Hence, the mixing regluation will also preserve localization and hyperuniformity properties. Though the mixing preservation of localization and global hyperuniformity is more notable, the fact that such a regulation barely affects the density fluctuations in the bulk is more general, which holds for even the non-hyperuniform cases. 

\section{Conclusions}
In conclusion, we show that demixing is absent for binary dimer (or rod-like) spinner mixtures due to a dynamical heterocoordination effect. Global hyperuniformity is also found to persist in this kind of mixed fluids, and can be enhanced or weakened by torques or densities in various ways. Correspondingly, a long wavelength scaling law $S(q\to 0)\sim  q^{\alpha}\,(\alpha>0)$ exists in the total structure factor of the system, where the exponent reaches a maximum value of about $1.8$ at around $\phi=0.25$. The deviation in the exponent from the theoretical value $2$ and the variation of the long-wavelength scaling law are attributed to the results of competition between the dynamical heterocoordination effect and the effective like-particle attractions. The absorbing transition with the critical scaling $S(q)\sim q^{0.45}$ is observed as well for this binary system, and the corresponding critical point is found to shift to an obviously higher density. When heated up by the driving torques of the spinners or through thermostating, the binary system exhibits a notable feature that all partial radial distribution functions become identical. This leads to a constant concentration-concentration structure factor, which prevents the system from being multi-hyperuniform.

As a potential application, such a hyperuniform mixing is further shown to be beneficial to the regulation of robust topological boundary flows. The great advantage of this mixing regulation that it barely affects the bulk density fluctuations even in a non-uniform flow field is rather intriguing. The regulation method also shows its delicacy in preserving localization and bulk hyperuniformity for systems with strongly-localized boundary flows. Experiments on rotors driven by light or electric/magnetic fields \cite{Zong2015,Zhang2022} may hopefully verify our results.

\section{Acknowledgement}
We are very grateful to Ran Ni for helpful discussions and comments. This work is supported by National Natural Science Foundation of China (Grant No. XXXXXXXX).

\end{document}